\documentclass[12pt,preprint]{aastex}

\slugcomment{ApJ. Draft: 09-Jul-2006}

\shorttitle{MWC 480 disk}
\shortauthors{Hamidouche et al.}

\begin{document}


\title{Resolving and probing the circumstellar disk of the Herbig Ae star MWC 480 at $\lambda$=1.4 mm: Evolved dust?}


\author{Murad Hamidouche, Leslie W. Looney}
\affil{Astronomy Department, University of Illinois, 1002 W Green st, Urbana, IL 61801}
\email{mhamidou@astro.uiuc.edu}

\author{Lee G. Mundy}
\affil{Astronomy Department, University of Maryland, College Park, MD 20742}



\begin{abstract}
We present high resolution 0.45\arcsec$\times$0.32\arcsec~ observations from the BIMA array toward the Herbig Ae system MWC 480 in the $\lambda$ = 1.4 mm dust continuum. We resolve a circumstellar disk of radius $\sim$170 AU and constrain the disk parameters by comparing the observations to flat disk models. These results show that the typical fit parameters of the disk, such as the mass, $M_D \sim$ 0.04-0.18 $M_\odot$, and the surface density power law index, $p$=0.5 or 1, are comparable to those of the lower mass T Tauri stars. The dust in the MWC 480 disk can be modeled as processed dust material ($\beta \approx$ 0.8), similar to the Herbig Ae star CQ Tau disk; the fitted disk parameters are also consistent with less-evolved dust ($\beta \approx$ 1.2). The possibility of grain growth in the MWC 480 circumstellar disk is supported by the acceptable fits with $\beta \approx$ 0.8. The surface density power-law profiles of p=0.5 and p=1 can be easily fit to the MWC 480 disk; however, a surface density power-law profile similiar to the minimum mass solar nebula model p=1.5 is ruled out at an 80$\%$ confidence level. 
\end{abstract}

\keywords{stars: individual: \objectname{MWC 480} --- stars: evolution, formation --- circumstellar matter --- techniques: interferometric
}

\section{Introduction}

Herbig AeBe stars (HAEBE) are young intermediate mass stars, ranging roughly from 2 to 20 M$_\odot$ \citep[]{her60}. They are the more massive counterparts of T Tauri (TT) stars and are usually optically visible and highly variable. They have broad emission lines and dust millimeter emission (e.g. Zuckermann 2001). Their pre-main-sequence evolution is more difficult to study with as much detail as TT stars because their formation and evolution processes are accelerated and more embedded; both lower mass Herbig Ae stars and the more massive Herbig Be stars are thought to accrete disk material quickly or disperse the disk by a stellar wind (e.g. Natta et al. 2000, N00 hereafter), which decreases disk lifetime. Nonetheless, there is increasing observational support for the existence of circumstellar disks around HAEBE stars (e.g. Fuente et al. 2003). This is interesting since the formation of planetesimals and planets is expected to occur in these disks during the pre-main-sequence evolution of the star. 

The principal objective in this study is to utilize subarcsecond observations of the Herbig Ae star MWC 480 circumstellar disk in the $\lambda$ = 1.4 mm continuum, probing its structure and morphology. MWC 480 (HD 31648) is a Herbig Ae star of spectral type A2/3ep and mass of 2.2 M$_{\odot}$ at a distance of 140 pc (Th\'e et al. 1994, Eisner et al. 2004). We will compare our results to previous observations of HAEBE stars and to their lower mass counterparts TT stars. We use a simple Gaussian model to fit these high resolution observations and estimate the disk parameters. Fitting a thin disk model with an inner hole to the observations allows us to deduce more explicitly the disk parameters. Constraining the disk is difficult with the large number of free parameters in the model. We fit the spectral energy distribution (SED) to constrain roughly the disk model. Our disk model fits the shape of the broad SED. This is a first step in our modeling strategy to reduce the number of free parameters. Thereafter, we use the allowed parameter space of the SED fitting to model the resolved disk image and better constrain the disk and best fit parameters. 

\section{Observations and data analysis}\label{OBS}

The observations were obtained with the Berkeley-Illinois-Maryland Association (BIMA) array\footnote{BIMA has since combined with the Owens Valley Radio Observatory millimeter interferometer, moved to a new higher site, and was commissioned as the Combined Array for Research in Millimeter Astronomy (CARMA) in 2006.}, located at Hat Creek, California. On 2004 February 9, MWC 480 was observed at $\lambda$ = 1.4 mm while the BIMA array was operated in its extended configuration (A array). The corresponding maximum baseline length was $\sim$ 1.2 km. System temperatures of the nine antennas varied from 436 to 975 K, and the weather was very good. We observed with a repeating cycle of 18.5 minutes on source and 3.5 minutes on the phase calibrator 0359+509. The amplitude calibrator MWC349 was observed at the beginning of the observations. Based on recent measurements, the assumed flux density of MWC349 during that period was 1.67 Jy, and the corresponding flux of 0359+509 was 2.67 Jy. The total observing time was 8 hours. On 2003 November 27, MWC 480 was observed with the B-array of BIMA. We used the same observing strategy and calibrators as in A-array. The assumed fluxes were respectively 1.67 Jy for MWC349 and 3 Jy for 0359+509. The system temperature ranged from 312 to 1100 K.    

The correlator was configured with a bandwidth per sideband of 625 MHz. The tracer line $^{13}$CO $J$=2$\rightarrow$1 was observed simultaneously with a resolution of 25 MHz. For mapping, we have combined A and B array data with robust = 0.1 weighting. The inclusion of B array data allows sensitivity to structures from about 0.35\arcsec~to 4\arcsec. The data were FFTed and CLEANed using the MIRIAD package software (Sault et al. 1995). The images were made separately for the lower and upper sidebands. We checked the consistency of the images then combined them to obtain the final image. The final beam size was 0.45$^{\prime\prime}$$\times$0.32$^{\prime\prime}$ at PA = 18\degr.

\section{Results}\label{Results} 

The continuum map at $\lambda$=1.4 mm of MWC 480 is shown in Figure 1. The emission arises from circumstellar dust around the star. The image resolves the disk that was previously detected by Mannings \& Sargent (1997, MS97 hereafter). This is the first resolved image of a dust disk around a Herbig Ae star at $\lambda$=1.4 mm. The total flux density of the circumstellar disk is 210$\pm$15 mJy and the peak value is 51$\pm$3 mJy/beam at R.A.(2000)=04$^h$58$^m$46$^s$.27 and Dec(2000)=+29\degr50$^{\prime}$36.95$^{\prime\prime}$. The error bars in the measurement represent only statistical uncertainties; we estimate an absolute flux calibration uncertainty of 20$\%$. Comparison with previous lower resolution interferometric observations at $\lambda$=1.4 mm shows a good agreement with the flux density 279$\pm$7 mJy (Mannings et al. 1997). This suggests that all emission is detected in our observations (cf. Natta et al. 1997). Lower mass TT stars show similar fluxes at $\lambda$=1.4 mm, in the hundreds of milliJanskys (Beckwith et al. 1990).

As a first step, we have fitted a simple Gaussian model and deduced a major axis $\sim$0.8\arcsec, minor axis $\sim$0.7\arcsec, and PA = 143$\pm$5\degr. Thus, the deduced source diameter (FWHM) is $\sim$100 AU at the source distance of 140 pc. The disk is inclined by 37$\pm$3\degr, which gives the elliptical shape. Both the disk PA and inclination show a good agreement with previous measurements: PA=148$\pm$1\degr and $i$=38$\pm$1\degr~from the $^{12}$CO $J$=2-1 emission (Simon et al. 2000). In addition to the disk emission, our image shows extended emission, total flux density $\sim$ 30 mJy, to the south-west region of the disk $\sim$0.7$^{\prime\prime}$ long ($\sim$100AU) at PA $\sim$ -160\degr. This may be due to a jet of free-free emission. Recent far-UV observations have detected a jet/counter-jet at a similar PA (C. Grady, 2005 private communication). 

At the resolution of Figure 1, we did not detect any $^{13}$CO $J$=2$\rightarrow$1 emission; presumably we resolve out any disk emission, and we are only sensitive to line brightness of 65 K (1$\sigma$ level) at this resolution, with a channel width of 5.3 km/s. In order to determine our sensitivity at large scales, we use the B array data with a beam-size of 1.5\arcsec$\times$1.3\arcsec. In that case, we have detected low-level $^{13}$CO $J$=2$\rightarrow$1 emission ($\sim 4\sigma$) that is offset by $\approx$ 1\arcsec~ toward the south-east. The flux density at the peak is 1.6$\pm$0.3 Jy/beam km/s within the velocity range 3.1 and 8.4 km/s, bracketing the V$_{lsr}$ of 5.6 km/s (Qi 2001). Qi (2001) detected blue-shifted and red-shifted line components of $^{13}$CO $J$=1$\rightarrow$0. The $^{13}$CO is important to trace the outer regions of the disk. It is mostly detected in larger circumstellar structures than our maximum detectable structures $\sim$ 560 AU. Simon et al. (2000) have reported the detection of $^{12}$CO $J$=2$\rightarrow$1 with a diameter $\sim$1100 AU. Even at our lower resolution map, we have still presumably resolved out much of the disk gas emission, and we are only sensitive to line brightness of 4 K (1$\sigma$ level), with a channel width of 5.3 km/s.

Assuming optically thin dust emission at millimeter wavelengths with a constant temperature, we can estimate the disk mass
\begin{equation}
M_{D} = \frac{F_{\nu} d^2}{\kappa_{\nu} B_{\nu}(T{_D})}  
\end{equation}
where F$_{\nu}$ is the flux, $d$ the distance to the source, and $\kappa_{\nu}$=$\kappa_0$($\nu$/1200$\mu$m)$^\beta$ the dust grain opacity (e.g. Beckwith \& Sargent 1991; Pollack et al. 1994). We adopt $\beta$=1 (e.g. Ossenkopf \& Henning 1994) and opacity coefficient $\kappa_0$ = 0.1 cm$^2$ g$^{-1}$ (Hildebrand 1983). In the Planck function B$_{\nu}$(T$_D$), we assume a constant disk temperature as suggested by N00 for an A3 star, T$_D$=23 K. We find a disk mass M$_{D}$ = 0.040 $M_{\odot}$ and a ratio of the disk to stellar mass $M_D$/$M_*$ = 0.018. Both of these values are within the ranges found by N00 for Herbig Ae stars.

Figure 2 shows the SED (Spectral Energy Distribution) of MWC 480, from the infrared to millimeter wavelengths, including our observation at $\lambda$ = 1.4 mm. The solid line is the SED obtained with power-law disk model \S \ref{SEDs}. A straight line fit to the measured dust emission yields a millimeter spectral index $\alpha$ $\simeq$ 2.48 (with $F_\lambda \propto \lambda^{-\alpha}$), within the wavelength range $\lambda$=0.45-2.7 mm. This index value is consistent with the typical spectral indexes for disks (Beckwith et al. 2000). 

Free-free emission contribution to the long wavelengths region of our SED is negligible (e.g. Rodmann et al. 2006). It is mostly dominant at centimeter wavelengths. Testi et al. (2003) deduced a contribution of the free-free emission in the Herbig Ae star CQ Tau of $\ll 10\%$ at 7 mm. 

\section{Disk Modeling}\label{model}   

In order to deduce the disk physical characteristics, we compare our observations to a flat disk model. We assume a flat, geometrically thin circumstellar disk with a hole in the center of radius $r_i$. The disk surface density is defined as $\Sigma$ = $\Sigma_0$ (r/1 AU)$^{-p}$ and the temperature $T$ = $T_0$ (r/1 AU)$^{-q}$, where $r$ is the disk radius, $T_0$ and $\Sigma_0$ are respectively the surface density and the temperature at the radius $r$ = 1 AU (e.g. Looney et al. 2000). Although this is a crude model, several studies have shown that circumstellar disks around Herbig Ae stars can be adequately characterized by a temperature power law and a surface density power-law (e.g. Testi et al. 2003). Note, we do not assume optically thin emission for this model.

Our modeling strategy is to make two complementary simulations. First, we model the SED in order to reduce the number of free parameters by fitting the shape of the broad observed SED. Second, we use the deduced parameters from this SED modeling for the simulation of the BIMA resolved disk image of MWC 480 at $\lambda$ = 1.4 mm.

\subsection{The Spectral Energy Distribution models}\label{SEDs}   

With the number of free parameters in the disk model, it is not possible to determine a unique model from the $\lambda$=1.4 mm data alone. Thus, we fix some parameters \textit{a priori}: $i$=37\degr~ and PA=145\degr, based on the results obtained in \S \ref{Results}. In addition, fitting the SED (Figure 2), that is the integrated flux at various wavelengths, will allow us to constrain the inner radius $r_i$, the temperature $T_0$ and its power-law index $q$, and the opacity index $\beta$, as these parameters affect the SED's shape and flux level. We leave the disk mass $M_D$ as a free parameter to optimize in the fit. As part of our modeling strategy, we have fixed $p$ at three discrete values 0.5, 1 and 1.5. Table 1 summarizes the parameter space explored in our SED models. 

\begin{center}
{\small Table 1. Range of variations of the disk model parameters.}       
\begin{tabular}{cccc}
\hline
\hline
Parameter & Lower limit & Upper limit & Steps \\
\hline
$R_o$ (AU) & 30 & 300 & 10 \\
        & 350 & 1250 & 100 \\
$r_i$ (AU) & 0.01 & 0.1 & 0.015 \\
        & 0.1 & 1 & 0.1 \\
$\beta$ & 0 & 2 & 0.01 \\
$T_0$ (K) & 160 & 370 & 10 \\
$q$ & 0.30 & 0.75 & 0.01 \\
\hline
\hline
\end{tabular}
\end{center}

We model the disk using all permutations of the parameters given in Table 1 (a total of $\sim 10^9$). For each disk model, we compute the SED and compare it to the observed SED, minimize the $\chi^2$ with respect to mass to select the best fit models. Likelihoods of at least 0.75 are considered acceptable in this SED modeling. 

Figure 3 (a) displays the likelihood obtained by varying $q$ and $\beta$ for each $p$ value. The other 3 most likely parameters were selected: the temperature $T_0$ at its best fit value $T_0$=290 K (Figure 3, b); the inner radius at $r_i$ = 0.046 AU ($\sim$ 5 stellar radii); and the outer radius for each $p$=0.5,1,1.5 at respectively $R_o$ = 150,200,300 AU (e.g. Table 2). We do not well constrain $R_o$ or $r_i$ in the SED modeling; to illustrate the other parameters, we chose representive values that are among the higher likelihoods of the SED modeling. The model disk mass $M_D$ and outer radius $R_o$ increase with $p$. This is due to the surface density power-law becoming steeper, hence more massive and larger outer radii disks are necessary to compensate and fit the emission at longer wavelengths (i.e. cold temperatures). Figure 3 (a) shows that the typical values are: $\beta\approx$1$\pm$0.2 for $p$=0.5 and $p$=1 and $\beta\approx$1.2$\pm$0.2 for $p$=1.5, and $q$=0.62 for the 3 values of $p$. The derived disk mass, $M_D$, increases with $\beta$; more mass is needed with a smaller opacity $\kappa_\nu$.   

Figure 3 (b) displays the likelihood obtained for varying $q$ and $T_0$. The outer disk radii were selected for each $p$=0.5,1,1.5 at respectively $R_o$ = 150,200,300 AU and inner radius $r_i$ = 0.046 AU. The opacity index is fixed to the best fit values deduced in panels (a); $\beta$=1 (for $p$=0.5 and 1) and $\beta$=1.2 (for $p$=1.5). The best fit models correspond to $T_0$ = 290 K and $q$=0.62, and they do not vary strongly with $p$. Moreover, displaying other likelihoods for other values of $\beta$ did not show a remarkable deviation from these best fit values of $T_0$ and $q$. 

Figure 2 shows the SED of a typical model that fits the observations. Table 2 summarizes the range of variation, upper and lower limits, of the disk parameters deduced in this SED modeling. The outer radius $R_o$ and $r_i$ are not well constrained by the SED alone. In addition, we could not constrain the surface density power law index $p$ with these SED fits. Modeling the higher resolution interferometric data better constrains $R_o$ and $p$ (see \S \ref{image}).

\begin{center}
{\small Table 2. Disk parameters deduced from best fit models to the SED. }
\smallskip
          
\begin{tabular}{ccccc}
\hline
\hline
$p$, $R_o$(AU) & $\beta$\tablenotemark{ a} & $r_i$(AU)\tablenotemark{ b} & $T_0$ (K)\tablenotemark{ c} & $q$\tablenotemark{ c} \\
\hline
0.5 , 150 & 0.67-1.47 & 0.01 - 0.08 & 270-310 & 0.59-0.67 \\
1 , 200  & 0.70-1.57 & 0.01 - 0.1 & 270-310 & 0.59-0.67 \\
1.5 , 300 & 0.75-1.77 & 0.01 - 0.1 & 280-310 & 0.59-0.67 \\
\hline
\hline
\end{tabular}\\ 
\medskip
{\tiny {\bf Note:} [a] Likelihood $>$0.75; [b] Likelihood $>$0.75 for $\beta \approx$ 1; [c] Likelihood $>$0.75 for $\beta$=1 (in $p$=0.5 and 1) and $\beta$=1.2 (in $p$=1.5).}
\end{center}

The most likely values found for $q$ are comparable to the values derived by Andrews \& Williams (2005) toward TT stars, using a flat disk model and assuming $\beta$=1. The flux density $F(r)\propto\Sigma(r) \times T(r)$ is proportional to $r^{-(p+q)}$, hence $q$ can be adjusted by changing $p$ since $p+q$ is not affected. The SEDs obtained are quite flat at shorter wavelengths, between 12-100 $\mu m$ (e.g. Figure 2). This is mostly due to the emission of the inner warmer part of the disk (within a few AU). The temperature index $q$ is constrained from this warmer inner region of the disk, hence at larger radii the disk model becomes less sensitive to $q$. 

\subsection{The 1.4 mm resolved image models}\label{image}   

In order to better constrain the disk model, we use our resolved observations, to further deduce disk parameters based on the SED parameter study \S \ref{SEDs}. The inner radius deduced from the SED is too small to be resolved in the millimeter image so it can not be fit; we fix $r_i$ = 0.046 AU. We fix the best fit values of $T_0$ = 290 K and $q$=0.62 and test models with different surface density indexes $p$ between 0.5 and 1.5. We fix $\beta$=1 in the case of $p$=0.5 and 1 and $\beta$=1.2 in the case of $p$=1.5. In practice, we fix $p$ and $\beta$, keeping the disk mass $M_D$ and the outer radius $R_o$ as free parameters to find the best fit model convolved with the beam. We determine the best fit by computing the residual image (the observation image minus the model image) then calculating the \textit{rms} of the noise $\sigma$, and the mean offset $\nu$ within the disk region. This region is defined by a rectangular box in the residual image that covers the location of the observed disk emission. The best fit is deduced by computing the flux density histogram averaged per beam size in this box and comparing it to a Gaussian function that represents the residual noise defined by $\sigma$ and $\nu$. Finally, we calculate the reduced $\chi^2_r$ level between the estimated histogram and the Gaussian function. We minimize $\chi^2_r$ and the mean offset $\nu$ in this region to match the noise by adjusting the mass. In addition, we fit the peak value of the flux density in two different maps made with the resolutions of 1\arcsec~ and 2.5\arcsec . This allows us to better constrain the disk mass.

\begin{center}
{\small Table 3. Range in fitted parameters obtained with a confidence level of 80$\%$}
\smallskip

\begin{tabular}{cccc}
\hline
\hline
$p$ & $\beta$ & $R_o$(AU) & $M_D$($M_\odot$) \\
\hline
0.5 & 1 & 130-210 & 0.041-0.07 \\
1.0 & 1 & 170-800 & 0.044-0.18 \\
1.5 & 1.2 & \small{no fits} & \small{no fits} \\
\hline
\hline
\end{tabular}
\end{center}

Table 3 summarizes the likely values of the deduced disk parameters. We did not find any combination of the parameters that fit the $p$=1.5 model at an 80$\%$ confidence level. We have constrained the outer disk radius $R_o$ for $p$=0.5 and 1. Although the deduced upper limit masses are relatively high compared to the expectations from the observations of Herbig stars (e.g. N00), these values can hardly be excluded as their likelihoods are high. On the other hand, we have found more models that well fit $p$=1 than $p$=0.5 for the parameters found in Table 3. This may be explained by the fact that for $p$=1 a higher disk mass can be fit compared to the $p$=0.5 models with a shallower surface density power law.

We summarize in Table 4 the typical fit models for each $p$ based on this image modeling. Typical fit models are displayed in Figure 4. One notices the increase of disk mass with $p$, as we have mentioned in \S \ref{SEDs}. For $p$=1.5, the mass increase is due to the increase of both $p$ and $\beta$.

\begin{center}
Table 4. Examples of fit models.
\smallskip

\begin{tabular}{ccccccc}
\hline
\hline
$p$ & $\beta$ & $R_o$(AU) & $M_D$($M_\odot$) & $\chi^2_r$ & Likelihood & $\sigma$(mJy/beam) \\
\hline
0.5 & 1 & 170 & 0.055 & 1.1 & 0.30 & 2.2 \\
1.0 & 1 & 250 & 0.061 & 0.9 & 0.47 & 2 \\
1.5 & 1.2 & 500 & 0.097 & 1.9 & 0.02 & 3 \\
\hline
\hline
\end{tabular}
\end{center}

Models of $p$=0.5 and $p$=1.0 fit the observed resolved image. In addition, the $rms$ levels corresponding to the disk region in the residual image of $p$=0.5 and $p$=1 models are both relatively consistent with the observed \textit{rms} noise $\sigma$ = 3.1 mJy/beam. The increase of the radius and mass with $p$, in these two models, can also be explained by the fact that the surface density profile becomes steeper and more mass at larger radii is necessary to compensate. The model shown in Table 4 for p=1.5 has only a $\approx 2\%$ likelihood, below our criterion for an acceptable fit. The corresponding SEDs of these models well fit the observations for the cases of $p$=0.5 and 1 (e.g. Figure 2) as well as for the case of $p$=1.5.  


We discussed only the examples among the models that fit the observations with the best fit values of $\beta$ deduced from the SED modeling (\S \ref{SEDs}). It is important to note that these fits, although having high likelihoods are not unique fits. Indeed, the $\beta$ values within the ranges given in Table 2 also provide good fits. These $\beta$ models fit both the 1.4 mm resolved image and the SEDs, as shown in the previous section. Furthermore, varying $\beta$ affects essentially the deduced disk mass in this image modeling at $\lambda$ = 1.4 mm, without statistically changing the derived parameters. For example, in the case of $p$=1: for $\beta$=0.8, we found a reduced $\chi_r^2 \sim$1 and a likelihood $\sim$0.41 with $R_o$=220 AU and $M_D \approx$0.04 $M_\odot$ and for $\beta$=1.2 we found a reduced $\chi_r^2 \sim$1 and a likelihood $\sim$0.41 with $R_o$=270 AU and $M_D \approx$0.09 $M_\odot$.


\section{Discussion} 

Both the SED and image fits using a \textit{geometrical thin} flat disk model are needed to adequately model the resolved $\lambda$=1.4 mm dust continuum observation of MWC 480. Modeling the SEDs alone provides a less compelling constraints on the disk parameters. Indeed, we have seen that some disk models' SEDs fit the observations, whereas their corresponding disk images do not fit the $\lambda$=1.4 mm observations. On the other hand, we have also found models that fit the disk image but not the SEDs. It is crucial to get subarcsecond resolution of the circumstellar structures in order to acquire more accuracy in constraining the disk parameters: the disk geometry and surface density profile. However, the upper limit of the disk outer radius is not as well constrained for the steeper surface density power laws. 

The infrared excess observed in the near-IR toward HAEBE stars has been a source of debate and has been interpreted by numerous authors as due to a flared component in the disk, where the dust is suspended and reprocesses the stellar optical light to longer wavelengths (e.g. Kenyon \& Hartmann 1987, Natta et al. 2001). However, at longer wavelengths the flared atmosphere dust emission is negligible (e.g. Hartmann et al. 1993, Natta et al. 2001). Thus, we have modeled only the midplane disk of the MWC 480 system. This feature affects its observed morphology and dominates in the far-IR and millimeter wavelengths. 


\subsection{Disk surface density} 

We determined from our $\lambda$=1.4 mm image and from the measured SED of MWC 480 that we can fit both $p$=0.5 and $p$=1. No models can be fit for the case of $p$=1.5 at an 80$\%$ confidence level. In the case of p=0.5, the likely disk mass and surface density power-law index $p$ are surprisingly comparable to the values of the TT star HL Tau (e.g. Mundy et al. 1996). In addition, the power law is also consistent with the observations of the zodiacal dust disk of our solar system p=0.34 (Mamajek et al. 2004). In the case of the p=1 model, we have obtained a higher mass and a larger outer radius $R_o$ (Table 3), compared to the p=0.5 model. These are due to the fact that the surface density power law is steeper and needs more mass and larger outer radius to compensate. Our typical disk masses are within the mass range given for Herbig Ae stars (N00). Our disk-star mass ratio $M_D$/$M_*$$\sim$0.02-0.07 is also within the range of Herbig Ae stars given by the same authors. Our deduced surface density distribution is shallower than the minimum mass solar nebula, $\Sigma \propto$ r$^{-1.5}$ (Weidenschilling 1977). Furthermore, Wetherill (1996) in his model formed planets from disks using both p=1 and p=1.5 profiles. Nevertheless, the surface density power-law can change due to the perturbations in the protoplanetary disk and become steeper (Raymond et al. 2005).


\subsection{Grain growth}

In both cases of $p$=0.5 and 1, we can exclude the possibility of the existence of only unprocessed dust (for which $\beta$=2.0) or only very large particles (e.g. large bodies, for which $\beta \rightarrow$0) in the MWC 480 disk, since both of these models do not fit the observations. A range of the opacity index, 0.8 $\leq \beta \leq$ 1.2, fit the observations with the $p$=0.5 and 1 values we have used. Although only one dust component may be present in the MWC 480 disk, it is probably a range of different size of processed particles (e.g. Pollack et al. 1994, D'Alessio et al. 2001). The opacity index $\beta$s are slightly higher in p=1 models than in p=0.5 (Table 2), which also explains the increase of mass as the opacity becomes lower. 


Our typical opacity indexes $\beta$=1.0$\pm$0.2 are consistent with grain growth into larger particles, millimeter-sized particles (e.g. Miyake \& Nakagawa 1993, Beckwith et al. 2000). These models do not exclude the existence of less processed particles in the disk corresponding to higher $\beta$ values. Testi et al. (2003) reported a similar value of $\beta \sim$0.6 using a flared disk model (Natta et al. 2001) to fit the resolved image of the Herbig Ae star CQ Tau at $\lambda$=7 mm. CQ Tau is less massive but older than MWC 480. Both these Herbig Ae stars have likely processed dust, and their disks may be considerably evolved at this age. 



\subsection{Possible remnant accretion process}

Outflows are usually considered to be linked to the accretion process of the stars (e.g. Fuente et al. 2001). The frequency of outflows toward TT stars is higher than their massive counterpart HAEBE stars, again confirming the faster evolution of HAEBE stars (Richer et al. 2000, Fuente et al. 2001). At this age, low mass stars are expected to be still accreting (e.g. Hartmann 1998), but as most of intermediate mass stars evolve faster than TT stars they probably have completed the accretion process at the age of MWC 480. The age of MWC 480 (a few Myrs) is comparable to classical TT stars (between Class II and Class III); Fuente et al. (1998) classified the HAEBE stars of this age as Type III not associated with outflows. However, if we confirm the detection of the jet-like emission, we could assume that there is a jet from the accretion process. Moreover, far-UV observations have shown a Herbig-Haro emission toward MWC 480 (C. Grady 2005, private communication; Williger et al. 2004) at PA=57.4$\pm$0.3\degr, i.e. the jet/counterjet are aligned with the disk minor axis (Figure 1).

\section{Conclusion}

We resolved the dust disk around the Herbig Ae star MWC 480 with the BIMA array at $\lambda$ = 1.4 mm. We fitted a geometrically thin flat disk model to our observed image and the SED, assuming no envelope or flared layers in the model. The best fit model consisted of a temperature power-law profile T(r)=T$_0$($r$/1AU)$^q$, with T$_0$=290 K and $q$=0.62. The typical opacity indexes are $\beta$=0.8-1.2. The highest $\beta$ values are mostly obtained for steeper surface density power law indexes $p$. We could not fit any model using $p$=1.5 profile of the standard minimum mass solar nebula model (Weidenschilling 1977). In the case of $p$=0.5 and $\beta$=1, the more likely fits included an outer disk radius $R_o$=130-210 AU and a disk mass of 0.041-0.068 M$_\odot$. Typical model parameters are: $R_o$=170 AU and M$_D$=0.055 M$_\odot$. In the case of $p$=1 and $\beta$=1, the more likely fits included an outer disk radius $R_o$=170-800 AU and a disk mass of 0.044-0.18 M$_\odot$. Typical model parameters are: $R_o$=250 AU and M$_D$=0.061 M$_\odot$. It is interesting to note that the surface density profile and disk mass found for the T Tauri star HL Tau (Mundy et al. 1996; Lay et al. 1997) are similar to our values for the Herbig Ae star MWC 480. We deduce that $\beta >$1 models, corresponding to less-processed grains, can not be excluded. We can only say that the grains may have grown into larger particles in the MWC 480 disk. This was already suggested for the TT star HL Tau (Lay et al. 1997). On the other hand, the extreme cases can be excluded by our models of: $\beta \rightarrow$ 0, for the existence of only large bodies in the disk and $\beta$=2 for the existence of only unprocessed dust primitively from MWC 480's parent molecular cloud or the ISM (Beckwith et al. 2000). 

From this study, we suggest that there has been some grain processing in the MWC 480 disk. There is an interesting resemblance between the two Herbig Ae stars MWC 480 and CQ Tau (Testi et al. 2003); both stars have similar opacity index $\beta$ and spectral index $\alpha$, and fit adequately a surface density power-law model with $p$=1. They may have evolved in the same manner. Finally, we have detected jet-like extention suggesting possible accretion. 

We show that both SED data and resolved images are necessary to constrain the circumstellar disk parameters. We have seen that we could accurately reproduce the flux densities (SEDs) without strong constraints on the morphological structure of the disk. In the near future, we will obtain more subarcsecond observations of Herbig AeBe stars allowing more constraints on the ensemble of circumstellar disk parameters to better understand their evolution. CARMA with its unprecedented sensitivity will be an excellent interferometer for these types of observations. 

\acknowledgments

We thank Yu-Shao Shiao for helping with the observations. M.H. and L.W.L. acknowledge support from the Laboratory for Astronomical Imaging at the University of Illinois and NSF AST 0228953. L.G.M. acknowledges support from NSF grant AST-0028963. L.W.L. and L.G.M. also aknowledge support from NASA Origin grant, NNG06GE41G and NNG06GE16G respectively.

\clearpage

\begin{figure}
\includegraphics[angle=-90,scale=.80]{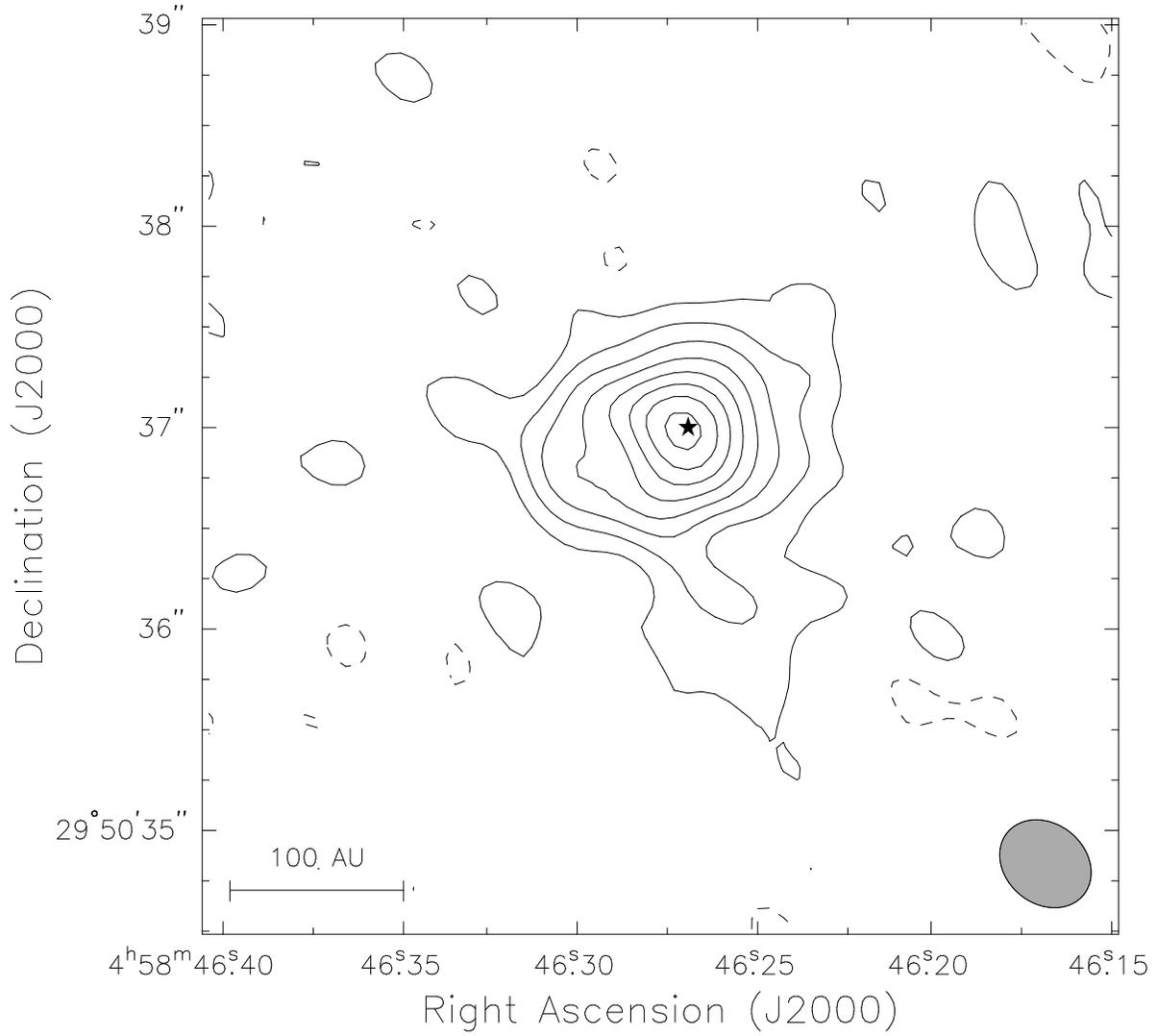}
\caption{MWC 480 in the $\lambda$ = 1.4 mm dust continuum. The image is made from BIMA observations with A and B array data combined with robust weighting of 0.1. Contours are in steps of (-4 -2 2 4 6 8 10 12 14 16)$\times$ $\sigma$ = 3.1 mJy/beam, where $\sigma$ is the RMS. The synthesized beam (0.45\arcsec $\times$ 0.32\arcsec, P.A. = 18\degr) is shown by the ellipse in the bottom-right corner. The star marks the optical position of the source (Perryman et al. 1997).\label{fig1}}
\end{figure}

\clearpage

\begin{figure}
\centering
\includegraphics[angle=-90,scale=0.65]{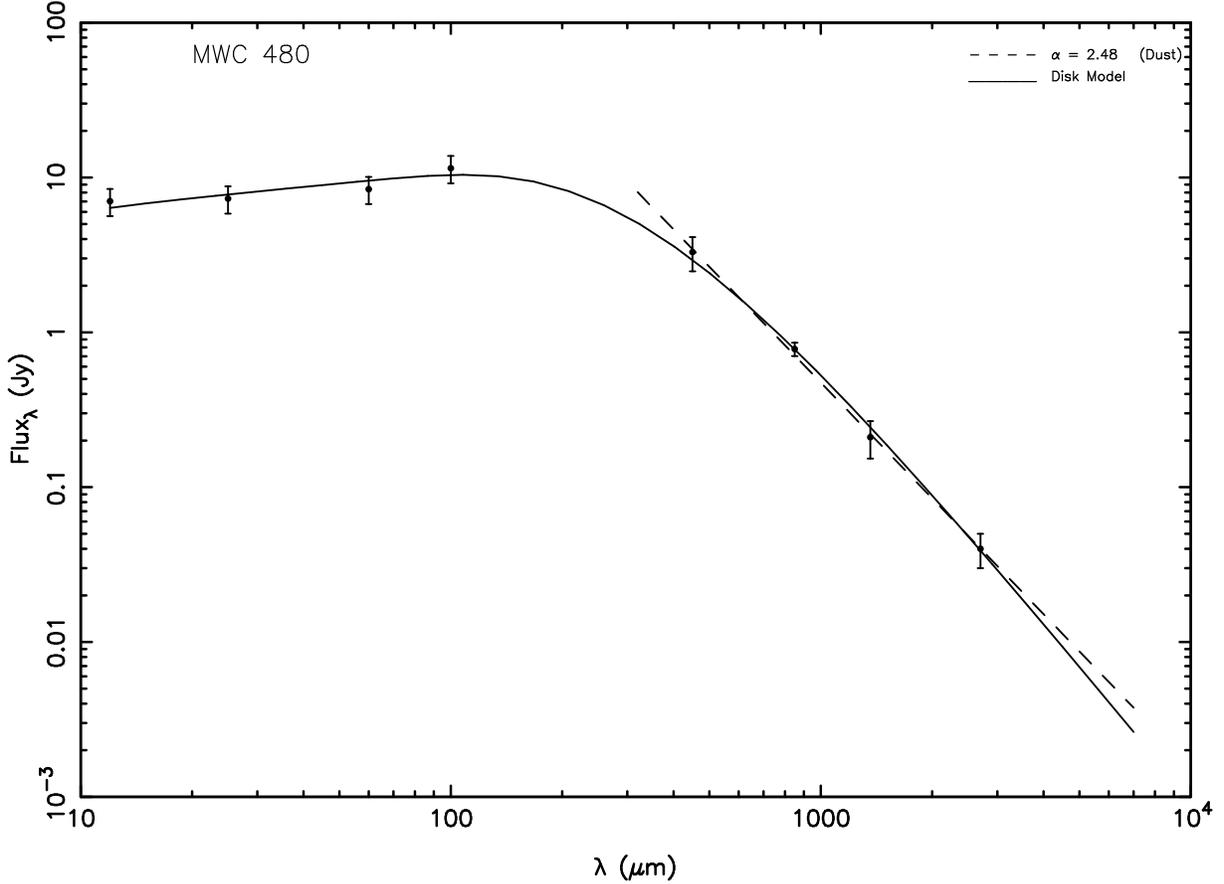}
\caption{Spectral energy distribution of MWC 480. The filled circles with error bars are the flux measurements. The four infrared fluxes $\lambda$=12-100 $\mu m$ are from IRAS (Creech-Eakman et al. 2002), the submillimeter fluxes at $\lambda$=450, 850 $\mu m$ are from the JCMT (Acke et al. 2004), the $\lambda$=1.4 mm measurement is our BIMA measurement, and the $\lambda$=2.7 mm is from OVRO (MS97). The dashed line is a fit of the observed fluxes of the circumstellar dust. It is characterized by a power law ($F_\lambda$ $\propto$ $\lambda^{-\alpha}$) where $\alpha$ = 2.48. The solid line is an example from a typical disk model. It well fits the observations within the error bars; corresponding likelihood $\approx$0.98. This is obtained with $T_0$=290, $q$=0.62, $p$=1, $M_D$=0.06$M_\odot$, $R_o$=250AU, $\beta$=1, and $r_i$=0.046 AU. \label{fig3}} 
\end{figure}

\clearpage

\begin{figure}
\centering
\includegraphics[angle=+90,scale=0.65]{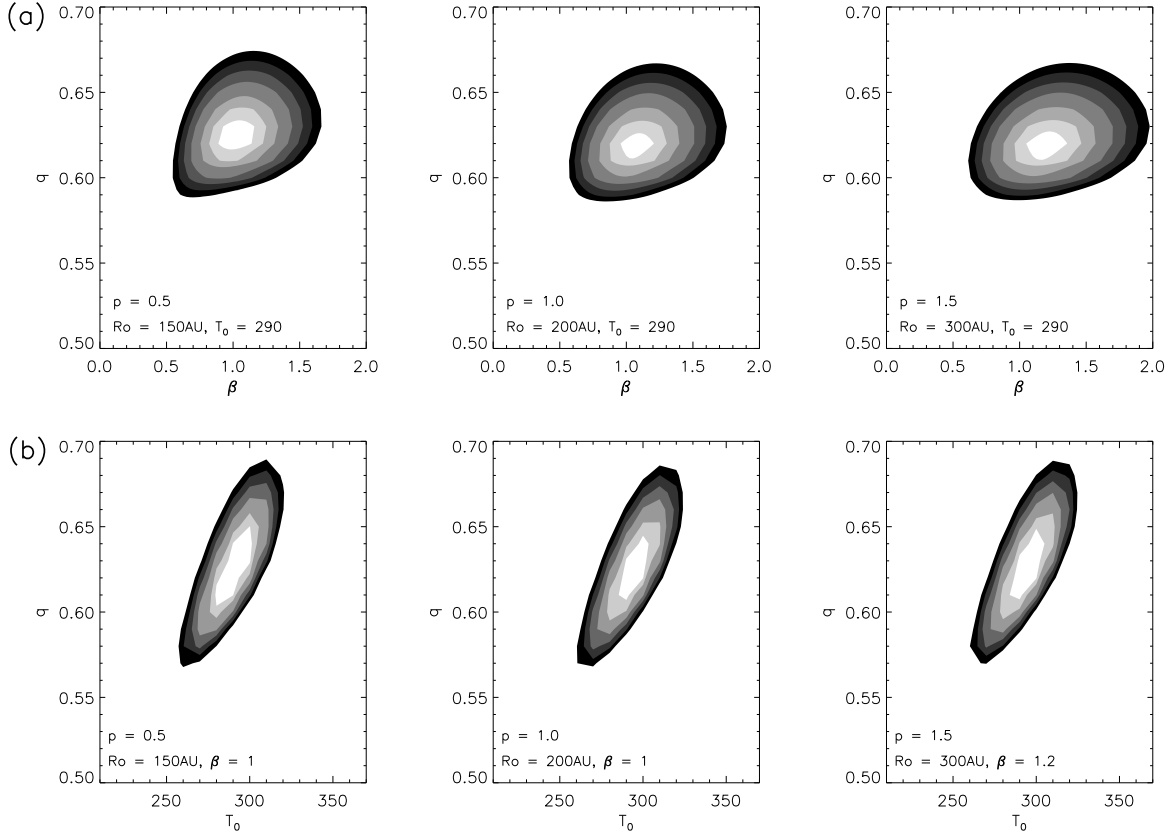}
\caption{Likelihoods from the comparison of the observed SED to power-law disk models' SED for different $p$ values. Panels (a) display the likelihood variation with $\beta$ and $q$ while $R_o$, $r_i$, and $T_0$ are fixed. The contour levels are 0.50, 0.60, 0.70, 0.80, 0.90, 0.95. Panels (b) display the likelihood variation with $T_0$ and $q$ while $R_o$, $r_i$, and $\beta$ are fixed. The contour levels are 0.50, 0.60, 0.70, 0.80, 0.90, 0.95, 0.975 \label{fig2}} 
\end{figure}

\clearpage

\begin{figure}
\centering
\includegraphics[angle=-90,scale=0.8]{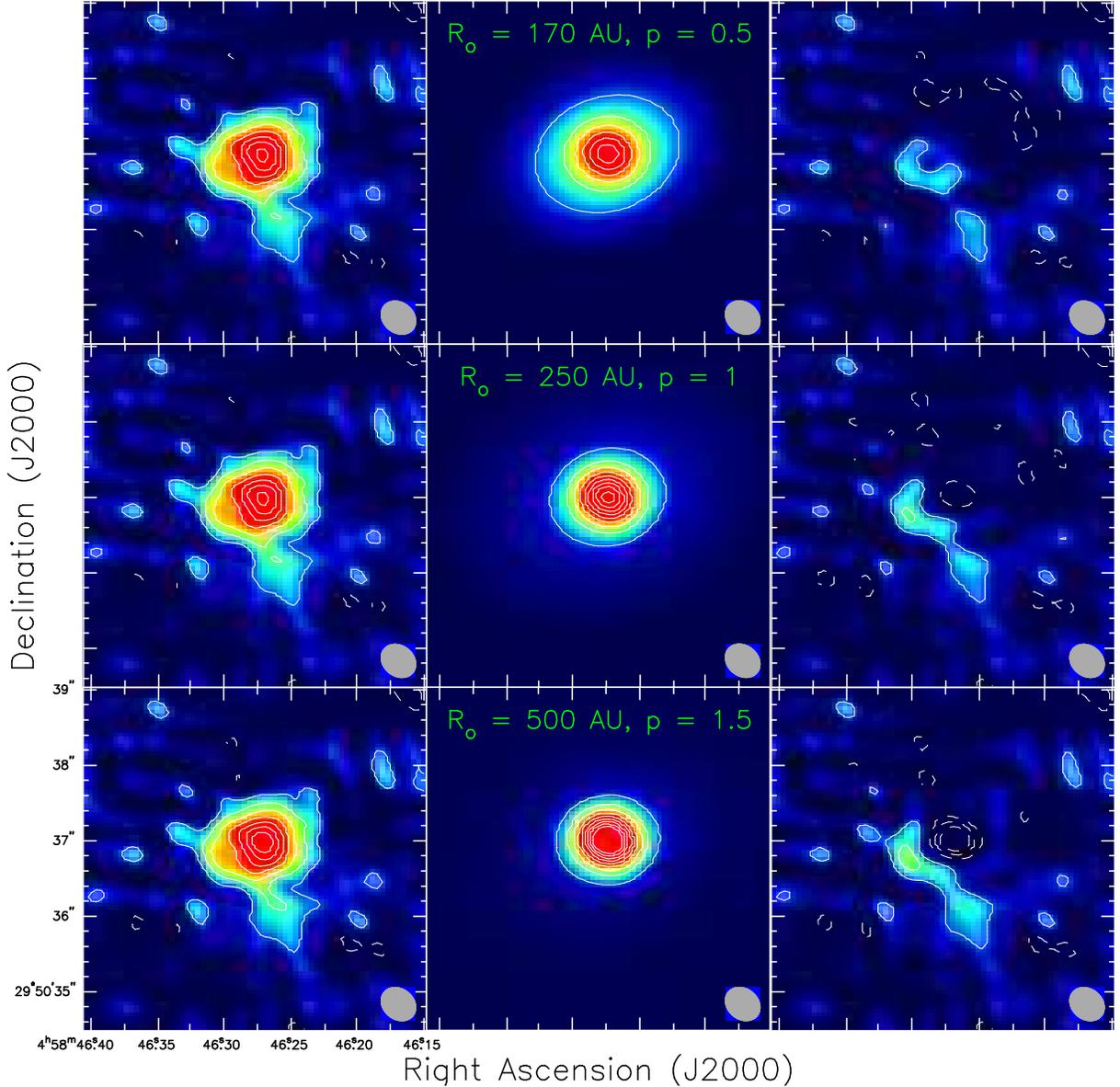}
\caption{The residuals (\textit{right panels}) in the image space obtained by subtracting the model convolved by the beam (\textit{middle panels}) from the observation of MWC 480 at 1.4 mm (\textit{left panels}). The parameters $q$=0.62 and $T_0$=290 K are fixed for all models. The panels correspond to $p$ = 0.5,1,1.5 respectively from the top to the bottom, with respectively $\beta$=1, 1 and 1.2. There are two free parameters values in each model respectively $R_o$ = 170, 250, and 500 AU and $M_D$ = 0.055, 0.061 and 0.097 $M_\odot$, see also Table 4 in the text. The contour levels are the same as in Figure 1. We chose the typical fit models based on their comparison with the observed image and their SEDs. The corresponding reduced $\chi^2_r$ levels are respectively from the top to the bottom $\approx$ 1.1, 0.9 and 1.9.\label{fig4}
}
\end{figure}

\end{document}